\documentclass{osa-article}
\journal{oe}
\articletype{Research Article}

\usepackage[utf8]{inputenc}

\usepackage{amsthm,bm}
\usepackage{todonotes}
\usepackage{amsmath,amsfonts}
\usepackage{tikz}
\usepackage{algorithm}
\usepackage[noend]{algpseudocode}
\usepackage{subfig}
\usepackage{sidecap}

\DeclareMathOperator*{\argmax}{arg\,max}
\begin{document}


\title{Adaptive Optics control using Model-Based Reinforcement Learning}
\author{Jalo Nousiainen\authormark{1,3} Chang Rajani\authormark{2}, Markus Kasper\authormark{3} and Tapio Helin\authormark{1}}

\address{\authormark{1} Department of Computational and Process Engineering, Lappeenranta--Lahti University of Technology, Finland\\
\authormark{2} Department of Computer Science, University of Helsinki, Finland\\
\authormark{3}European Southern Observatory, Karl-Schwarzschild-Str. 2, 85748 Garching bei München, Germany\\
}

\email{\authormark{1}jalo.nousiainen@lut.fi} 

\begin{abstract}
Reinforcement Learning (RL) presents a new approach for controlling Adaptive Optics (AO) systems for Astronomy. It promises to effectively cope with some aspects often hampering AO performance such as temporal delay or calibration errors. We formulate the AO control loop as a model-based RL problem (MBRL) and apply it in numerical simulations to a simple Shack-Hartmann Sensor (SHS) based AO system with 24 resolution elements across the aperture. The simulations show that MBRL controlled AO predicts the temporal evolution of turbulence and adjusts to mis-registration between deformable mirror and SHS which is a typical calibration issue in AO. The method learns continuously on timescales of some seconds and is therefore capable of automatically adjusting to changing conditions.
\end{abstract}

\section{Introduction}

Atmospheric turbulence distorts astronomical imagery obtained with ground-based telescopes. Adaptive optics (AO) \cite{babcock1953possibility, hardy1998adaptive, roddier1999adaptive} is a technique that aims at minimizing the distortions caused by the turbulence. In AO, a wavefront emitted by an astronomical object, such as a star, and distorted by the atmosphere is directed to one or more deformable mirrors (DM) before it propagates to the scientific camera. The distortions are measured with a wavefront sensor (WFS), and optimal image quality is obtained by setting the DM to a shape that partially cancels the distortions after reflection. In this work, we consider the classical single conjugated AO (SCAO) system, which requires a bright star that is close to an object of interest. This reference star is used to calculate distortion caused by the atmosphere along the propagation path. Since the atmosphere is continuously evolving, the mirror's shape has to be controlled in real-time, often from 300 to more than 1000 times a second.

\begin{figure}[t]
\centering
    \includegraphics[width=\textwidth]{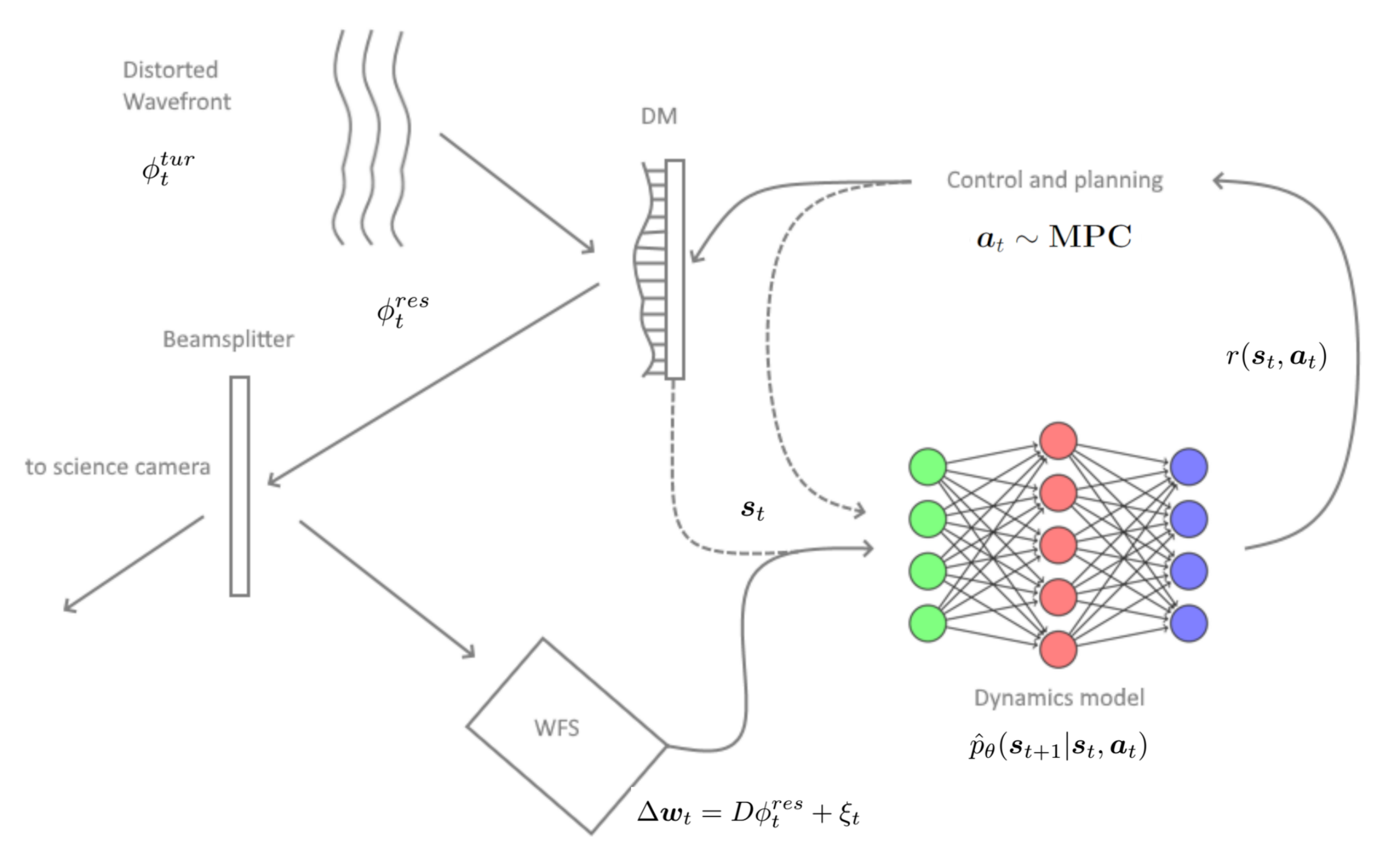}
    \caption{Overview of the task and method. The distorted wavefront is propagated into the deformable mirror (DM), which is controlled by our control algorithm. The algorithm reads the wavefront sensor (WFS) input, simulates how it will evolve using the learned dynamics model, and plans for the next DM commands with a process called model predictive control (MPC).}
    \label{fig:methods_mbrl}
\end{figure}

Most AO systems run in a closed-loop configuration, where the WFS measures the wavefront distortions after DM correction; see Figure \ref{fig:methods_mbrl}. The goal of such a control-loop is to minimize the distortions in the measured wavefront i.e., the residual wavefront. For high contrast imaging (HCI) the wavefront error budget (within the AO controlled region) is often dominated by the temporal delay error \cite{guyon2005limits}. Also real systems often suffer from a dynamic mis-alignment between DM and WFS called mis-registration \cite{heritier2018new}. Reinforcement learning (RL) provides an automated approach for control, which promises to cope with these limitations of current AO systems. Unlike the classical control methods, RL methods aim to learn a successful closed-loop control strategy via interacting with the system. Hence they do not require accurate models of the components in the control loop and adapt to a changing environment.

In recent years, the merger of RL and deep neural networks (NN), called deep RL, has become increasingly popular due to its effectiveness in problems with large state- and action-spaces. This type of RL has been used, for example, to play video- and board-games on a superhuman level \cite{mnih2013playing, silver2017mastering} and for vision-based real-world robot control \cite{zhang2015towards, kalashnikov2018qt}. Much of the success can be specifically attributed to \emph{model-based} RL (MBRL), where a model of the environment is learned using data obtained by interaction, and a \emph{planning algorithm} is used in conjunction to decide the next action. Inspired by these successes, we attempt to generalize the adaptive optics problem to the general framework of reinforcement learning and apply existing algorithms in solving it.

Our starting point is to formulate the closed-loop AO system as a Markov decision process (MDP), the prevailing mathematical framework for reinforcement learning \cite{sutton2018reinforcement}. We describe the state of the AO system as a finite time series of past control voltages and WFS measurements and assume that such a state exhibits Markovian statistics to a good approximation, i.e. each state depends only on the previous state, where a state can also include data from several timesteps from the past. The key to successful prediction lies in finding a reliable model for the system dynamics. Here, we parameterize the dynamics model describing the conditional distribution of the next state given the current state and action using standard NN architectures. This parameterization is fitted to closed loop data in a process called training. Using this framework, we adapt a standard state-of-the-art MBRL algorithm, Probabilistic Ensemble Trajectory Sampling (PETS) \cite{chua2018deep}, to train the model and optimize for the next action, i.e. the set of control voltages. 

The paper's structure is as follows: In Section 2, we state the novelty of our method and position our work with respect to existing literature. In Sections 3, we give a short description of an AO control loop and the baseline method. Section 4 describes MDP formulation of the AO control loop, setting a platform for RL. Further, we describe the algorithm used and how we adapt it to AO. For small details and a general more in-depth justification of the method, the authors strongly encourage the reader to have a look at the original paper on the algorithm \cite{chua2018deep}. In Section 5, we demonstrate the performance of our method, through simulation of a small and simple SCAO system controlled either by RL or by the baseline integrator controller.  The algorithm and MDP formulation presented is only one way to solve the control loop with RL but already hints at the great potential of MBRL control for AO. Finding the optimal formulation and bringing the computational time of the controller to the required level are left for future research. Finally, Section 6. discusses the topic, especially how MBRL could be implemented in a real system and how to overcome the significant hurdle of inference time and computational jitter.

\section{Related work}
In order to mitigate the measurement noise and temporal error, predictive controller methods have been proposed for ground-based adaptive optics. These methods include the Kalman filter-based linear quadratic Gaussian control (LQG) \cite{kulcsar2006optimal, paschall1993linear} and its variants \cite{gray2012ensemble, conan1a2011integral, correia2010adapting,correia2010optimal, correia2017modeling} and predictive filters operating on separate modal coefficients such as Zernike polynomials or Fourier modes \cite{poyneer2007fourier, males2018ground, dessenne1998optimization}, which provide up to a factor of 1000 gain in raw point spread function contrast in an idealized simulation environment for an extremely large telescope at very small angular separations and using a very bright AO guide star \cite{males2018ground}. The contrast performance gets shallower for larger angular separation, smaller telescopes or fainter stars. More recently, data-based predictive methods have emerged in AO literature. Examples include linear predictive filter methods such as empirical orthogonal functions \cite{guyon2017adaptive}, the low-order linear minimum mean square error predictor \cite{van2017performance, van2019impact, haffert2021datadriven}, as well as NN-based methods \cite{swanson2018wavefront,liu2019using, sun2017bayesian, mcguire1999adaptive}. 

Many of the existing machine learning-based predictive control methods \cite{swanson2018wavefront, sun2017bayesian, liu2019using} have not been studied in  a closed-loop configuration, but in principle, they can be integrated into closed-loop systems by utilizing a so-called pseudo-open loop telemetry \cite{guyon2017adaptive, jensen2019demonstrating}. These procedures consist roughly of two steps: collecting open-loop wavefront estimates from pseudo-open loop telemetry and learning a predictive filter as a supervised learning task. This procedure assumes accurate knowledge of system time lags and DM response, as well as close to linear behavior of the WFS. As a consequence, the predictive filter will inherit the errors in system calibration. These methods, therefore, learn the temporal evolution of the turbulence from the data but rely on modelling of the system components and interactions between them, which leads to the need for external tuning and re-calibration of the predictive controller to ensure robustness. Moreover, some AO systems operate at framerates which are high enough that usually neglected system dynamics (e.g. the finite response time of the DM) become important. Consequently, a control algorithm suffers from the simplifying assumption of a temporal step-wise response.

In contrast to these methods, we present a technique that learns predictive and noise-robust control straight from the system feedback without the set of prior assumptions mentioned earlier and eliminating the need for accurate calibration or modeling assumptions. Our RL formulation uses a generic Neural Network (NN) architecture to build the dynamics model. NNs have been applied to various aspects of AO before. The topics vary from open-loop systems to the extraction of Zernike coefficients directly from the images and to non-linear wavefront reconstruction; see \cite{gomez2019experience,osborn2014open, gonzalez2017comparative, sandler1991use, Landman:20}.

Also RL-based concepts have already been applied to AO. Self-adaptive control has been studied in \cite{xu2019deep}, where a deep learning control model is proposed to mitigate alignment errors in the calibration. Model-free RL methods for wavefront sensorless AO have been studied in \cite{ke2019self, hu2018build}, where the method is compared against stochastic parallel gradient descent providing improved correction speed. Finally, model-free RL for ground-based AO was implemented to control tip and tilt only \cite{landman2020self}. The model-free RL method they used learns a policy NN that directly outputs the two values for the tip and tilt mirror given the observations. Such methods often require a large number of interactions with the environment, which increase exponentially with the degrees of freedom to be controlled if no additional measures are taken. In contrast, we control each actuator of a high-order DM via model-based RL, formulate ground-based astronomical AO as a general MBRL task, and discuss its potential benefits. We show that state-of-the-art model-based RL learns a self-calibrating noise-robust predictive control law using only a few seconds of past telemetry data.

\section{Adaptive optics and the classical integrator}
\label{sec:baseline}
We first present the adaptive optics task, along with useful notation, and then frame it in the reinforcement learning setting. An overview of the AO control loop is given in Figure \ref{fig:methods_mbrl}. The incoming light $\phi_t^{tur}$ at the timestep $t$ gets corrected by the DM. After this correction the WFS measures the residual wavefront $\phi_t^{res}$. Commonly, a linear relationship between the WFS observation and the residual wavefront is assumed, i.e.,  
\begin{equation}
    \Delta \bm w^t = D \phi_t^{res} + \xi_t,    
\end{equation}
where $\Delta \bm w^t = (\delta w_1^t,\delta w_2^t, \cdots, \delta w_n^t)$ is the WFS data and $D$ is so-called interaction matrix modelling the WFS measurement and $\xi_t$ is the measurement noise typically composed of photon and detector noise. Depending on the type of WFS, a component $\delta w_i$ of the residual wavefront can represent, e.g., a wavefront modal coefficient, a wavefront slope or the wavefront phase itself. Classical control algorithms are often modelled by a linear mapping of the WFS measurements $\Delta \bm w$ to the residual DM control voltages $\Delta \bm v$ i.e.,
\begin{equation}
    \label{eq:dm_to_wfs}
    \Delta \bm v^t = C \Delta \bm w^t,    
\end{equation}
where $C$ is so-called reconstruction matrix. To obtain the reconstruction matrix, we decompose the DM on a Karhunen--Loeve (K-L) modal basis. Each mode of the K-L basis has a representation in terms of actuator voltages. This relation is fully determined by a linear map $B$ from voltages to modes.  The $B$ matrix is computed by a double diagonalization process, which takes into account the geometrical and statistical properties of the telescope \cite{gendron1994astronomical}. In the following, we utilize a reconstruction matrix defined by the Moore--Penrose pseudo-inverse 
\begin{equation}
C = (DB)^{+}.
\end{equation}
We truncate the number of K-L modes in B to have a stable inversion and a reasonably low noise amplification by C.

Let us now consider a simple non-predictive control algorithm known as the \emph{integrator law}. At a given timestep $t$, the WFS measures the residual wavefront. The new control voltages $\Tilde{\bm v}^t$ are obtained from
\begin{equation}
\Tilde{\bm v}^t = \Tilde{\bm v}^{t-1} + g C \Delta \bm w^t,
\end{equation}
where $g$ is the integrator gain. In order to stabilize the loop, the value of $g$ is often fixed below a value of about 0.5 for a two-step delay system \cite{madec1999control}. Large values of $g$ increase the correction bandwidth, i.e., the loop reacts faster. On the other hand, a large gain reduces the control loop's stability margin and amplifies noise propagation. The challenge in classical integrator control is in balancing these two effects to minimize the average error of the method \cite{gendron1994astronomical}.

In the following we denote the vector concatenating the past $m$ control voltages

\begin{equation}
\label{eq:past_voltages}
\bm V^m(t) =  ( \Tilde{\bm v}^{t-1},   \Tilde{\bm v}^{t-2} \hdots  \Tilde{\bm v}^{t-m} )^\top,
\end{equation}
and the vector concatenating the past $k$ residual voltages, constructed from the WFS slopes, by
\begin{equation}
\label{eq:residual_voltages}
\Delta \bm V^k(t) =  (  C \Delta \bm w^{t-1},    C\Delta \bm w^{t-2}, \hdots,   C\Delta \bm w^{t-k} )^\top.
\end{equation}
This quantity merely represents WFS measurements in the voltage space projected on the K-L modal basis defined by B. It does not represent voltages applied to the DM.

On the millisecond time scale of AO operations a big part of turbulence is presumably in frozen flow and the turbulence evolution is predictable to some extend \cite{poyneer2009experimental}. Control methods that use past telemetry data have shown a great potential both in turbulence prediction and noise reduction
\cite{guyon2017adaptive}. In a closed-loop set-up, these methods would, for example, utilize past control and residual voltages in equations \eqref{eq:past_voltages} and \eqref{eq:residual_voltages}, respectively, to construct a pseudo-open loop data stream used for the prediction. This paper aims to obtain a controller with similar properties but without the need for neither an accurate knowledge of time delay, accurate calibration nor a linear response of the WFS to wavefront errors.

\section{Adaptive Optics as Model-based Reinforcement Learning}
\label{sec:mdp}

\subsection{Markov decision process and the dynamics model}
We model the closed-loop adaptive optics control problem as an MDP. An MDP consists of a set of states $\mathcal{S}$, a set of actions $\mathcal{A}(s)$ at the given state $s$, a set of transition probabilities $p(s_{t+1}|s_t,a_t)$ and a reward function $r(s_t, a_t)$. 

In AO, the set of actions consists of different combinations of control voltages, and the state consists of the prevailing atmospheric turbulence and the shape of the mirror during the measurement. In practice, we do not have access to the full state of the AO system, i.e., full turbulence, wind speeds and DM shape. We only partially observe the state through a noisy WFS measurement. Consequently, past observations and actions are still valid information for the prediction of the next observation. To account for partial observation and to ensure the Markovian property of state formulation, we define the state as a sequence of previous voltages and residual voltages derived from WFS measurements:

\begin{equation}
\bm s_t =  \begin{pmatrix} \bm V^m(t) \\  \Delta \bm V^k(t) \end{pmatrix},
\end{equation}
where we typically choose $k = m$. The state includes data from the previous m (or k) time steps and the reconstruction matrix $C$. We stress that the residual voltages are not applied to the DM. They are merely a quantity closely related to the residual wavefront through eq. \ref{eq:residual_voltages}, and which the MBRL control approach (see Section \ref{sec:pets}) will try to minimize. The matrix $C$ must only be chosen such that the residual voltages are well observable by the WFS. It does not have to match the actual registration of DM and WFS precisely and could be given by either a previous calibration or derived from a system model. Moreover, previous studies have shown that the neural network-based wavefront reconstructor benefits from involving a linear control matrix with a non-linear WFS \cite{Landman:20}, and we observe below that MBRL is robust to errors or perturbations in the reconstruction matrix; see Section \ref{misreg}.

The action of the MDP is simply a vector of the changes to the control voltages 
\begin{equation}
\bm a_t = \Delta \Tilde{\bm v}^{t}.
\end{equation}

Let us now represent the true transition probability $p(\bm s_{t+1}| \bm s_t,\bm a_t)$, i.e. the conditional distribution of the next state (including the next WFS residual) given the current state and action as a parameterized distribution family $\hat{p}_{\theta}(\bm s_{t+1}|\bm s_t, \bm a_t)$. The aim of MBRL is to find the optimal approximative model $\hat{p}_{\theta}$ given a data set from the real environment. We solve this problem by fitting NNs using straightforward supervised learning, detailed in Section \ref{sec:pets}. In our case, the parameters $\theta$ represent the weights of the neural networks. The transition probability approximations  $\hat{p}_{\theta}$ represent our probabilistic dynamics and are hence called the dynamics model. It provides an estimation of the next state (of which only the next WFS measurements are new) from the current state and the control voltages. The dynamics model involves information about the interaction of voltages with WFS measurement as well as the system's temporal evolution, including the turbulent wavefront.

In adaptive optics we aim to minimize the residual wavefront $\varphi^{\text{res}}$ over the the whole time interval. The most natural reward for an AO system would be the Strehl ratio, or for a high contrast imaging (HCI) instrument the contrast obtained. Since we are considering a control system with just one WFS, we can only choose a reward function observable on that specific sensor. We choose a reward for a state-action pair as the residual voltages' negative squared norm corresponding to the next measurement: 
\begin{equation}
r(\bm s_t, \bm a_t) = -\| C \Delta \bm w^{t+1}\|^2.  
\end{equation}
This quantity is proportional to the observable part of the negative norm of the true residual wavefront. The WFS measurement is blind to some modes, e.g., the waffle mode for a Shack-Hartmann Sensor (SHS). We ensure that we do not control these modes by projecting each action, i.e., set of control voltages to the control space. That is,
\begin{equation}
\bm a_t = B^{+}B\Delta\Tilde{\bm v}^{t},    
\end{equation}
where $B^{+}B$ projects the control voltages onto the control space defined by the K-L modes.  

\subsection{Model-based reinforcement learning}\label{MBRL}
Now that we have defined the MDP components and the dynamics model, we can outline our MBRL approach. First, we initialize an empty data set, and we initialize the dynamics model parameters $\theta$ (the weights of the NN) randomly from a zero-mean Gaussian distribution. Then, we collect our first data set by running the AO loop for a particular time interval (an episode) with random actions (DM control voltages) sampled from a zero-mean Gaussian distribution as well.

After the first episode we have the first data set and use it to train the dynamics model. The training is described in more detail in Section \ref{sec:pets}. 

We now have a first reasonable guess for the dynamics model and start to use it during the second episode to find the action that maximizes the expected future reward (minimizes the residual voltages) for a given state. This optimization task is called \emph{planning} and replaces the regulator/controller in classical AO. We detail the methods used for this in Section \ref{sec:cem}.

After the second and subsequent episodes, the previous data set is concatenated with the new data and the dynamics model is trained again and updated. When the data set gets sufficiently long, old data is removed to ensure that the NNs are trained on sufficiently fresh data only. The dynamics model is entirely learned from data obtained while running the loop, i.e., during the experiment; no simulation or modeling steps are involved.

\subsection{The PETS algorithm}
\label{sec:pets}
We implement the MBRL control for AO approach described above following the PETS algorithm \cite{chua2018deep}. 
We use OOMAO \cite{conan2014object} to simulate the AO system plant (turbulence, telescope, DM, WFS), and the Probabilistic Ensemble Trajectory Sampling (PETS) algorithm replaces the classical reconstruction, control and calibration. The algorithm combines a probabilistic ensemble (PE) neural network dynamics model and model predictive control (MPC) \cite{camacho2013model} that is based on trajectory sampling (TS). We combine the TS with the cross-entropy method (CEM) as described in Section \ref{sec:cem}.

\subsubsection{The dynamics model}\label{ensemble}
Our choice of the dynamics model, an ensemble of probabilistic NNs, can model two types of uncertainty. Firstly, it models the uncertainty associated with the predictions, e.g., the stochastic behavior of the turbulence and measurement noise, by outputting a variance estimate in addition to a mean prediction. Secondly, it models the uncertainty associated with the model's parameters by learning an ensemble of bootstrap models. Each model has its unique data set to be trained upon that is bootstrap sampled (a statistics term meaning sampling with replacement) from the whole data set recorded so far\cite{efron1994introduction, lakshminarayanan2016simple}.

In preparation for the experiment, we verified that using an ensemble of NNs leads to a superior correction performance as a single NN. Then, we also ran tests and confirmed that estimating the next state's variance improves the performance compared to a fixed variance. Both measures combined stabilize training by a fair amount and eventually reach a higher reward, i.e., a better correction performance.

Each neural network in the ensemble defines a parameterized distribution family $\hat{p}_{\theta}(\bm s_{t+1}|\bm s_t, \bm a_t)$ satisfying
\begin{equation}
    \label{eq:param_model}
    \hat{p}_\theta(s_{t+1} | s_t, a_t) \sim \mathcal{N}(\mu_\theta(s_t, a_t), \sigma^2_\theta(s_t, a_t)),
\end{equation}
where the mean $\mu_\theta(s_t, a_t)$ and the variance $\sigma^2_\theta(s_t, a_t)$ of the Gaussian field are outputs of a neural network. We train the dynamics model ensemble by maximizing the log-likelihood of a Gaussian for which the parameters are outputs of the neural network model. More specifically, given a dataset of $N$ transitions $\mathcal{D} = \{(s_t^i, a_t^i), s_{t+1}^i \}_{i=1}^N$ we maximize the following objective function 
\begin{equation}
\label{eq:objective}
    \hat \theta = \argmax_\theta \log \prod_{i=1}^N \hat{p}_\theta(s_{t+1}^i | s_t^i, a_t^i) 
\end{equation}
where $\hat{p}_\theta$ is given by equation \eqref{eq:param_model}. Each network that is a part of the ensemble is trained similarly, but with different bootstrap sampled data set from $\mathcal{D}$. Each network is modelled as a convolutional neural network with 2 hidden layers of 8 feature maps each. Both layers are activated by a leaky rectified linear unit (LReLU) \cite{Maas2013RectifierNI}. We use the concatenated vector $[s_t, a_t]$ as an input and output the mean and log-scale variance of a normal distribution: the distribution of the next state. The maximization in equation \eqref{eq:objective} is done using an extension of stochastic gradient descent called the Adam algorithm \cite{kingma2014adam}. The neural network hyperparameters (e.g., number of layers, convolutional features maps, activation function used) provided relatively fast implementation and performed well in our experiments. We did not tune them further because of the large number of hyperparameters and that the method was not very sensitive to them. However, moving to more complex numeric simulations or lab experiments, hyperparameters have to be more extensively studied. A full pseudocode is given in Algorithm \ref{alg:pets}, where $\varnothing$ stands for empty set and $\mathcal{D} \gets \mathcal{D} \cup \mathcal{D}^{(\text{new})}$ for concatenation of previous dataset and new data set that was collected during the last episode.

\begin{algorithm}

\caption{PETS for Adaptive Optics}
\label{alg:pets}
\begin{algorithmic}[1]

\Function{PETS}{}
    \State Initialize dataset $\mathcal{D} \gets \varnothing$
    
    \For{episode in $1 \dots$}
        \State Initialize dynamics model $\hat{p}_\theta$ randomly
        \State Train $\hat{p}_\theta$ on $\mathcal{D}$ for $L$ epochs using Eq. \eqref{eq:objective}
        \State Record transitions $\mathcal{D}^{(\text{new})} = (s_t, s_{t+1}, a_t, r_t), t \in 1\dots T$ by running $\text{CEM}(s_0, \hat{p}_\theta)$ in simulator for $T$ timesteps
        \State Set $\mathcal{D} \gets \mathcal{D} \cup \mathcal{D}^{(\text{new})}$
    \EndFor
\EndFunction

\end{algorithmic}
\end{algorithm}

\begin{algorithm}

\caption{Cross-Entropy Method (CEM) for planning in AO}
\label{alg:cem}
\begin{algorithmic}[1]

\Procedure{CEM}{$s', \hat{p}_\theta$}
    \State $\mu \gets a_{t-1}, \sigma^2 \gets \sigma^2_0$
    
    \For{i in $1 \dots n_{\text{iters}}$}
        \State Sample actions $a_{t,\dots,t+2} \sim \mathcal{N}(\mu, \sigma^2_0)$
        \State Set $s_t \gets s'$
        \For{t in $1 \dots T=2$}
            \State Sample possible next states $s_{t+1} \sim \hat{p}_\theta(s_t, a_t)$
            \State Observe rewards $r(s_t, a_t)$
        \EndFor

        \State Select elites $\hat{a}_1, \dots, \hat{a}_{n_{\text{elites}}}$ corresponding the largest rewards
        \State Update $\mu \gets \text{mean}(\hat{a})$ and $\sigma^2 \gets \text{Var}(\hat{a})$
        
    \EndFor
    
    \Return $\mu$
    
\EndProcedure

\end{algorithmic}
\end{algorithm}

\subsubsection{Planning control}
\label{sec:cem}

We use the learned dynamics model to plan for the action, i.e., the mirror commands to apply at each timestep. The goal of the planning algorithm is to optimize a sequence of actions $\{a_t, a_{t+1} \cdots a_{t+T}\}$ such that it maximizes the expected reward inside some planning horizon $T$ \cite{camacho2013model}.

For the AO case, the action $a_t$ taken at timestep $t$ takes one timestep to be executed, and one additional timestep for the corresponding observation to be recorded. Therefore, we are essentially doing planning to minimize the observed wavefront sensor measurements up to $s_{t+2}$, i.e., we implicitly predict the best control action by the DM at the time of the WFS measurement (two frames into the future in this case). This planning horizon of two steps provides stable control to time delays smaller or equal to 2 frames.  On a real AO system the time delay is to some extend stochastic and/or non integer. Therefore, the planning horizon should include the longest time delays that may occur in the control loop. Further, in the presence of DM dynamics the effective planning horizon might be a couple of time steps longer, since the control voltage decision are not fully independent.

Starting at the given initial state, the CEM works as follows. We first sample a trajectory of actions $a_t, a_{t+1}$ from a Gaussian distribution parameterized by some starting $\mu$ and $\sigma^2$. Next we use the learned dynamics model $\hat{p}_\theta$ to produce a sequence of potential next states given the actions and the initial state, i.e., $s_{t+2} \sim \hat{p}_\theta(s_{t+1}, a_{t+1})$, where $s_{t+1} \sim \hat{p}_\theta(s_{t}, a_{t})$. Since the dynamics model is approximated by an ensemble, these states will include samples trained using different bootstrapped training datasets. The algorithm then chooses the so-called \emph{elites}: actions that produce the best rewards, and recalculates the sampling distribution parameters $\mu, \sigma^2$ to adjust to the elites using a maximum likelihood estimate. Finally, the mean of the sampling distribution is returned as the best trajectory. Note that in the actual task only the first action is executed, after which another transition is observed, and the algorithm is run again using the new observation as the starting state. This procedure of re-planning at each timestep is often referred as model predictive control (MPC). The full pseudo-code is given in Algorithm \ref{alg:cem}.
 

\label{sec:results}

\begin{table}[ht]
    \centering
    \caption{}
    \label{table:simulator_parameters1}
\begin{tabular}{ |p{1.2cm}|p{1.1cm}|p{1.7cm}|p{1.7cm}|p{1cm}|p{1.8cm}| } 
 \hline
 \multicolumn{6}{|c|}{Atmospheric turbulence layers} \\
 \hline
         & $C_N^2$ ($\%$)  & speed ($m/s$) & direction ($^{\circ}$) & $L_0$ ($m$) & altitude (km) \\
 \hline
 Layer 1   &  70  & $15$   &   $0$  & 30 & 0 \\
 Layer 2   &  25  & $3$   &   $45$  & 30 & 4\\
 Layer 3   &  5   & $7.5$   &   $90$  & 30 & 10 \\
 \hline
  \multicolumn{6}{|c|}{Misregistration parameters} \\
 \hline
         & -  & shift ($\%$)  & direction ($^{\circ}$) & - & - \\
 \hline
 Case 1   &  -  & $14$  &   $225$  & - & - \\
 Case 2   &  -  & $28$   &   $135$  & - & - \\
 \hline
\end{tabular}

\end{table}

\section{Results}

\subsection{Simulation set-up}
In the following numerical simulations, the OOMAO simulator serves as the plant of the control system - it only provides the WFS measurements and receives a vector of the control voltages. The PETS algorithm runs in Python and interacts with the plant via Python/MATLAB interface.

We compare the results against the ones obtained by a well-tuned integrator controller as well as a theoretical controller that suffers neither from time delay nor measurement noise. This theoretical controller is computed from the non-delayed noiseless measurement, i.e., it still contains errors due to the aliasing and uncontrolled high order modes. The same limitation also applies to the MBRL and integrator controllers. The optimum integrator gain is always tuned globally to give the best performance (Strehl ratio) at each simulation set-up (GS magnitude and misregistration (MR)) separately. This is done manually, and typical values were between $0.3-0.6$ for our simulation setups.

\begin{figure}[ht]
\centering
    \includegraphics[trim={2cm 0 2.2cm 0}, clip,width=0.9\textwidth]{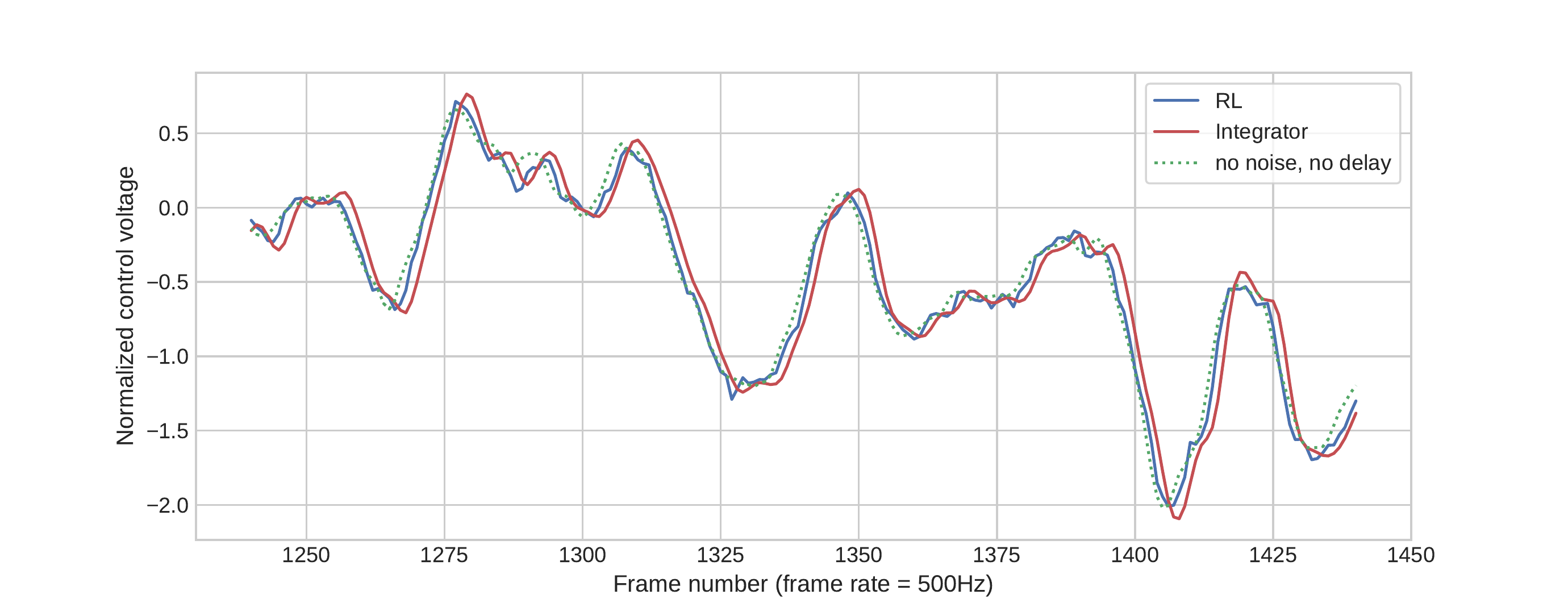}
    \caption{Predictive control at low-noise (0th mag ngs) regime. We plot a short section of the control voltage time series during the evaluation of each control method: the RL method (blue), the integrator (red), and the theoretical limit of having no noise nor delay (dashed green line). Here we see that the integrator suffers from the time delay, whereas the RL method closely follows the non-delayed signal.}
    \label{fig:timeseries}
\end{figure}

\begin{figure}
\centering
    \includegraphics[width=0.8\textwidth]{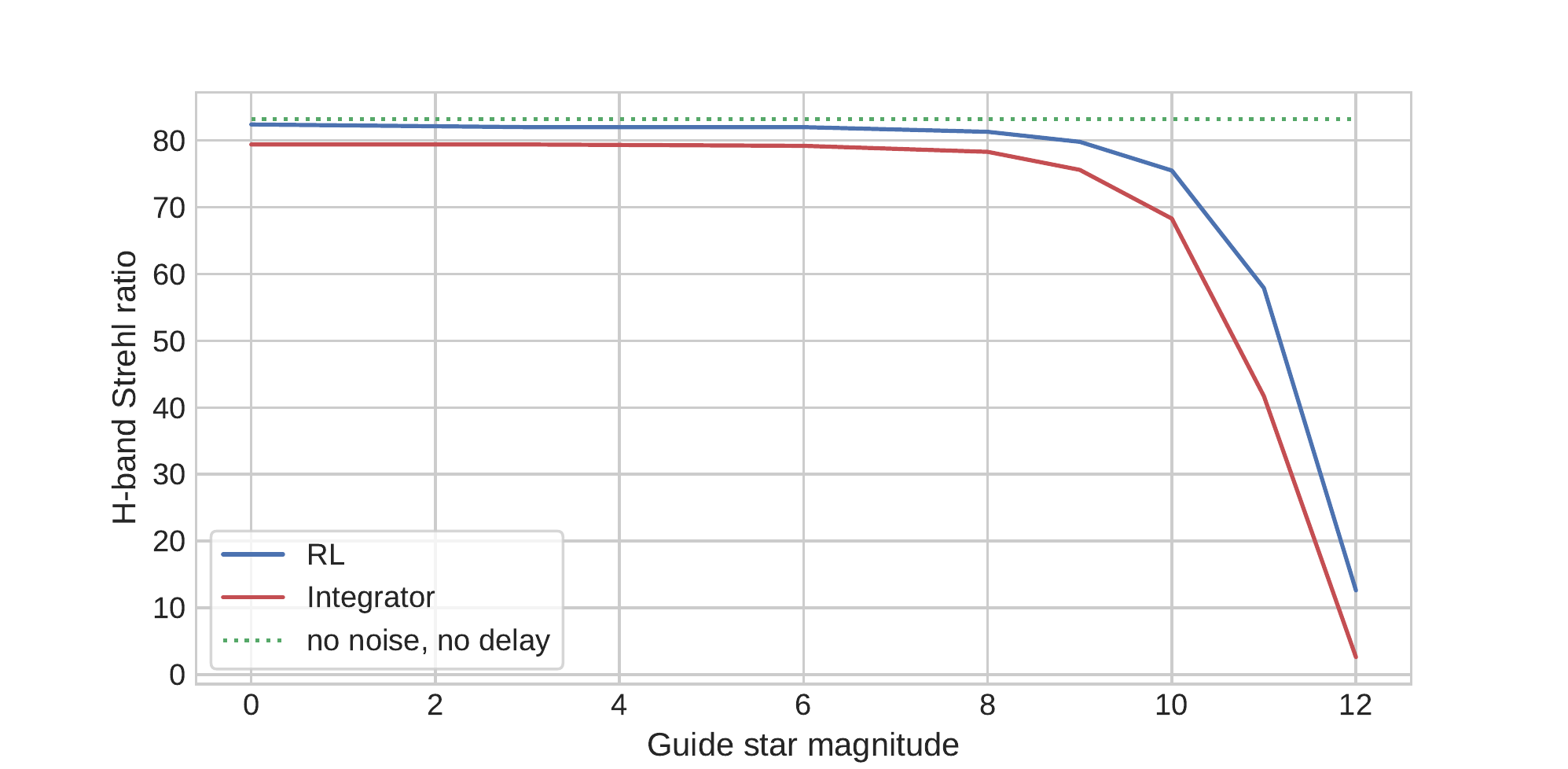}
    \caption{Results summary. A comparison of H-band Strehl ratios (SR), with respect to star magnitude (GS). The red lines correspond to adapted Integrator and the blue lines the RL method. The dotted green line is the theoretical limit of having no noise nor delay (dashed green line). The RL algorithm always outperforms the integrator and is close to the theoretical limit in the low noise regime.}
    \label{fig:strehl}
\end{figure}

We simulated an $8$m telescope observing a single natural guide star (NGS), equipped with a $23 \times 23$ SHS, and a $24 \times 24$ DM with a Fried geometry (actuators on the subaperture corners). The DM actuator influence functions are assumed to be Gaussian with a 45\% coupling. Atmospheric turbulence is simulated as a sum of three frozen flow layers with Von Karman power spectra combining a Fried parameter $r_0$ of $15$ cm at $550$ nm wavelength. The parameters of the atmosphere are listed in Table \ref{table:simulator_parameters1}. The loop is running at a framerate of 500 Hz with a time delay of $2$ steps. We pick the simulation parameters to demonstrate three key properties of the proposed method:

\begin{itemize}
    \item The predictive capacity of the method is shown on a system with a negligible measurement noise; see Figures \ref{fig:contrast1} and \ref{fig:timeseries}, and Table \ref{table:simulator_parameters2}.
    \item The robustness of the method against observation noise is shown by observing natural guide stars of different magnitudes; see Figures \ref{fig:strehl}, \ref{fig:contrast_all} and Table \ref{table:simulator_parameters2}.
    \item The self-calibrating property is demonstrated by running the same simulations but introducing (MR) between the WFS and DM; see Figure \ref{fig:contrast4} and  see Table \ref{table:simulator_parameters1}. 
\end{itemize}

We model MR in calibration by changing the alignment between the WFS and the DM in two different directions and shift amplitudes (see Table \ref{table:simulator_parameters1}). All images and contrast plots are calculated at $\lambda = 1.65\mu m$ (H-band), and the WFS measures at $\lambda = 551 nm$  (V-band). Wind speed and MR are somewhat pessimistic to prevent the error budget from being dominated by the significant aliasing error of the SHS \cite{rigaut1998analytical}. 

\begin{table}[ht]
    \centering
    \caption{}
    \label{table:simulator_parameters2}
\begin{tabular}{ |p{5cm}|p{2.2cm}|p{3.2cm}| } 
 \hline
 \multicolumn{3}{|c|}{Simulation parameters} \\
 \hline
         Parameter  & value  &  Units  \\
 \hline
 Telescope Diameter   &  8  & m     \\
 Obstruction ratio    &  14  & percent           \\
 Sampling frequency   &  500  & Hz        \\
 Active actuators     &  448   & actuators         \\
 WFS subapertures     & $23\times23$ & apertures           \\
 WFS pixels           & $10 \times 10$ & pixels                \\
 WFS diffraction limited FWHM & $2 \times 2$ & pixels   \\
 read-out noise     & 5    & photo-events rms       \\
 Photon flux 0/9/10 mag  &   2033k /511/ 210    &   photons / frame / lenslet          \\ 

  \hline
 \multicolumn{3}{|c|}{MPC parameters} \\
 \hline
         Parameter  & value  &  Units  \\
 \hline
 Planning horizon (T)     &  2  & steps     \\
 Past DM commands (m)     &  4  & commands       \\
 Past WFS measurements (k)  & 4  & frames        \\
 CEM elites/particles    & 200/2000     &                \\
 CEM iterations            & 20  &                       \\
 PETS ensemble size        & 3   &              \\             
  \hline
\end{tabular}

\end{table}

We set the state of the MDP $s_t$ to include the last four actions and four WFS measurements and set the episode length to 400 frames giving a balance between a fast iteration and a reliable performance estimate. We validate our proposed algorithm by running multiple simulations in the simulator. Each simulation starts with the knowledge of the reconstruction matrix, but zero knowledge of temporal behavior including the time lag. Note here that the sole purpose of the reconstruction matrix is to implement the control space filtering by mapping WFS measurements on residual voltages to be included in the state. We never change it when running the MBRL control, in particular we do not update it to match the MR. Our model learns to compensate for the measurement noise, misregistration in the reconstruction matrix, and the atmosphere's temporal behavior by interacting with the environment.

\begin{figure}[ht]
\centering
    \includegraphics[trim={2.2cm 0 2cm 0},
    width=\textwidth]{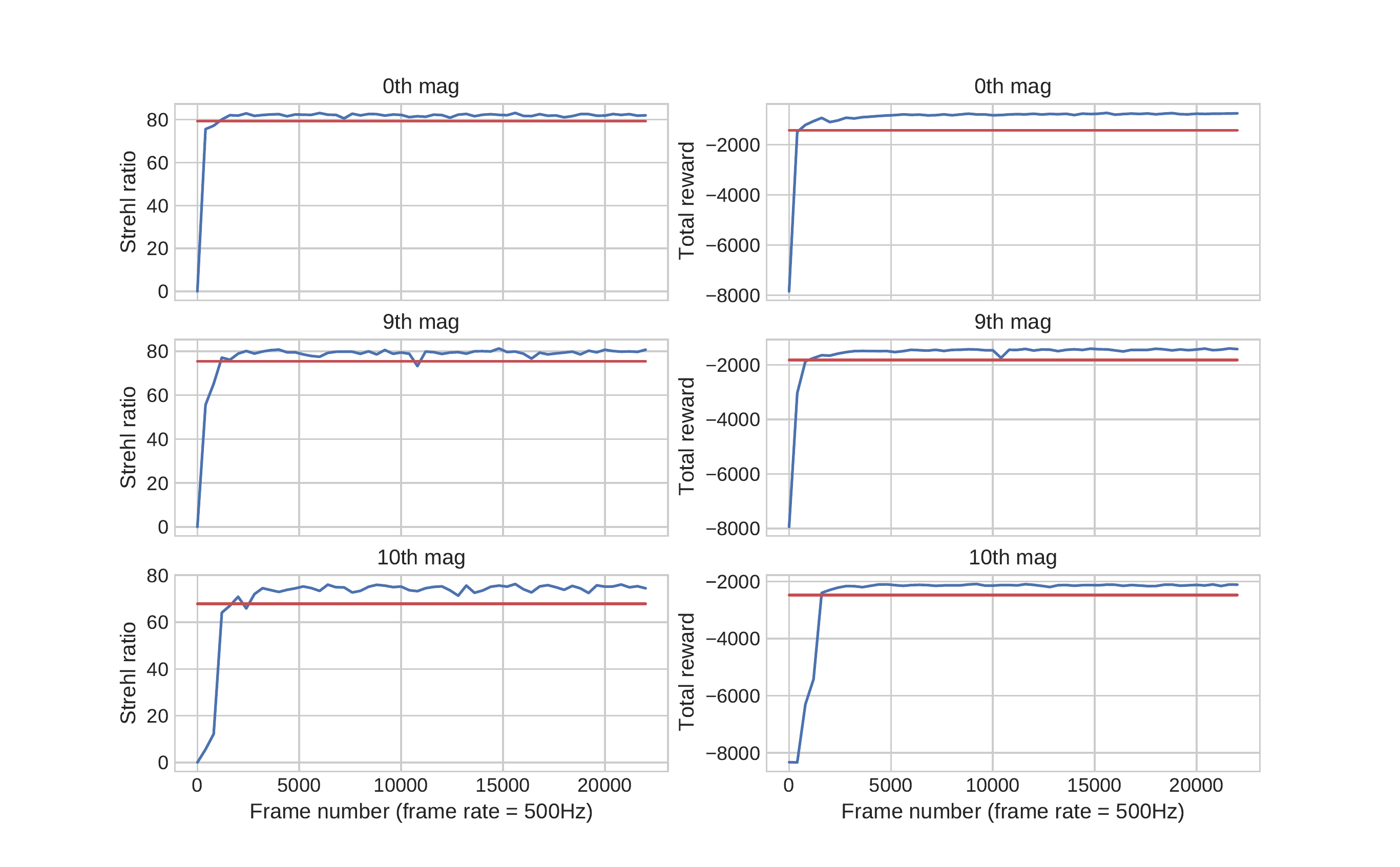}
    \caption{Learning curves for the proposed RL method. Each learning curve represents the episode performance in H-band Strehl ratios (left) or total reward (right), defined as the sum of rewards at each frame. The red lines are the mean performance of the integrator (Section \ref{sec:baseline}) and blue lines the performance of PETS after each episode. The learning process itself tries to maximize the total reward and the corresponding improvement in Strehl ratios is a consequence of this process. Our model converges in around 10 episodes, or 4000 frames. The performance level of the integrator control is passed already in after 4 episodes, i.e., 1600 frames.}
    \label{fig:results_training}
\end{figure}

\subsection{Training}
To demonstrate how fast the method learns a successful control strategy in different noise conditions and MRs, we compare the learning curve of the method to the baseline of the integrator; see Figures \ref{fig:results_training} and \ref{fig:results_training_mr}. In terms of loss i.e., the negative reward over the episode, our model outperforms the integrator baseline after about 1600 frames and reaches its full potential in about 4000 frames, in all of the test cases. The total loss in the figure corresponds to the sum of normalized residual voltages computed from the WFS measurements. For the simulated system running at 500 Hz, 1600 timesteps is equivalent to 3.2 seconds of actual time, while 4000 is 8 seconds. As described in Section \ref{MBRL}, we train and update the dynamics model after each episode. The loop is suspended during this time, which amounts for a several seconds given our rather shallow NN architecture and moderate computational power. At the telescope with typically variable observing conditions (wind speed and directions, seeing, guide star magnitudes), the dynamics model has to be trained in parallel to the observation, for example using a separate computer. The available time for training is then set by the episode length and should not exceed the time-scale of environment variability.

\subsection{Prediction and noise robustness}
\begin{figure}
\centering
\subfloat[\label{fig:contrast1}]{%
  \includegraphics[trim={4.5cm .1cm 2cm 1.15cm}, clip,width=0.9\textwidth]{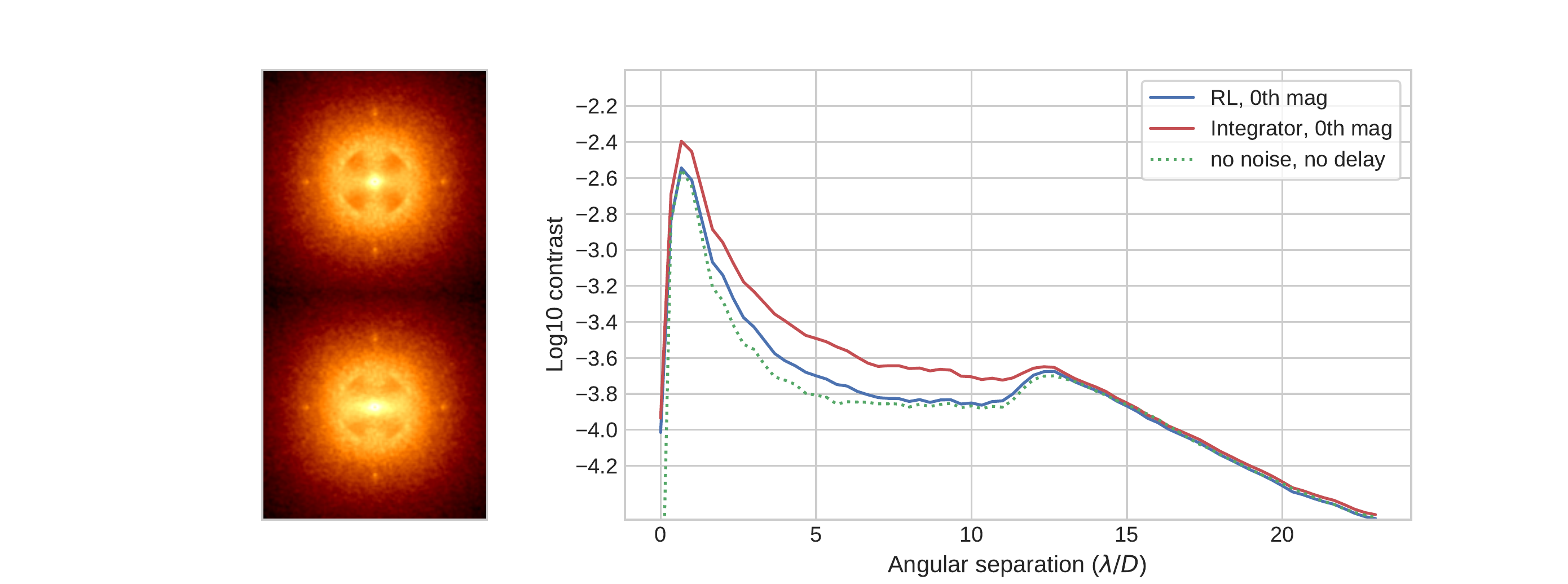}%
  }\par
\subfloat[\label{fig:contrast2}]{%
   \includegraphics[trim={4.5cm .1cm 2cm 1.15cm}, clip,width=0.9\textwidth]{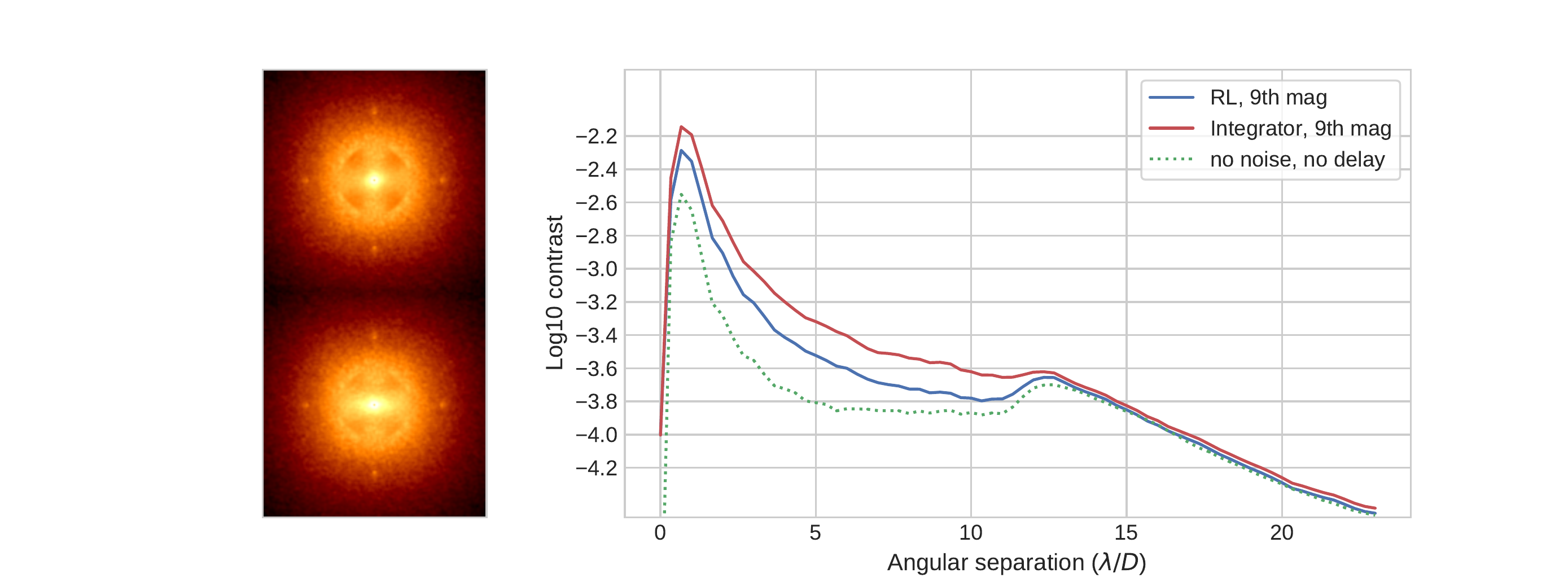}%
  }\par        
\subfloat[\label{fig:contrast3}]{%
  \includegraphics[trim={4.5cm .1cm 2cm 1.15cm}, clip,width=0.9\textwidth]{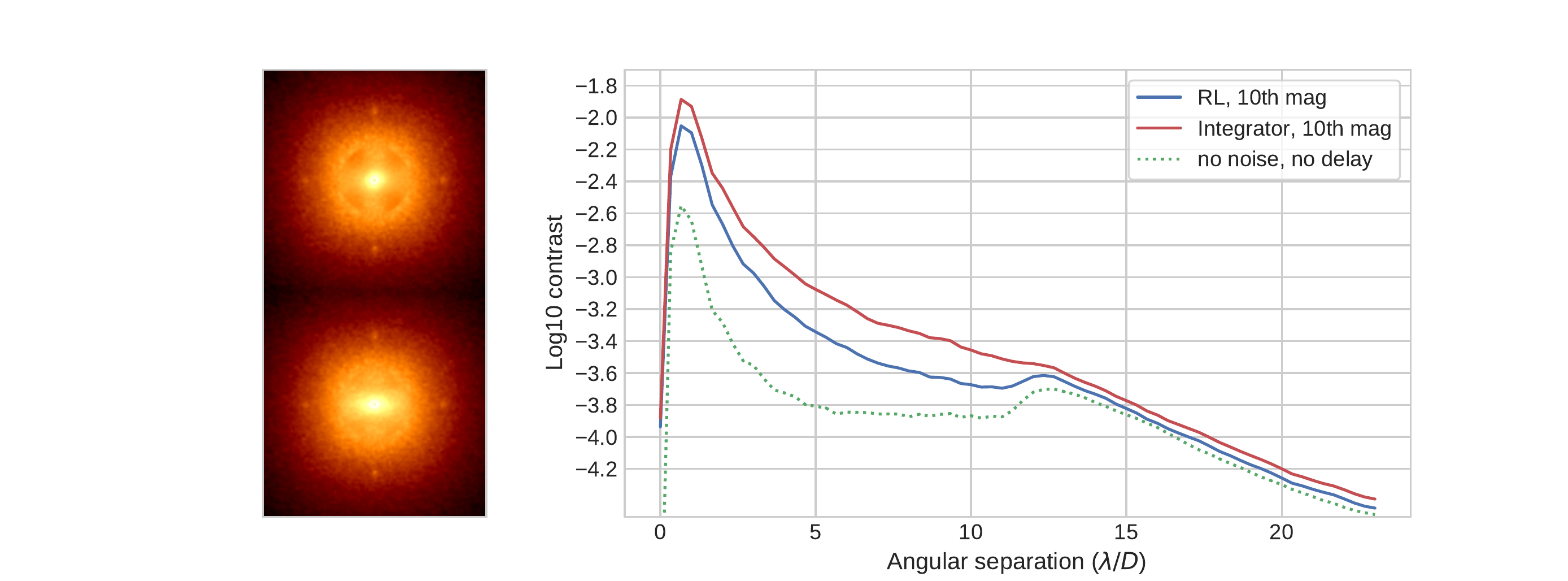}%
  }
  
\caption{Contrast benefit on three different noise levels. Left: Raw PSF contrast on the pupil plane for RL method (upper panel) and Integrator (lower panel). Right: azimuthal average of the images. The blue lines are for RL method and red for the integrator. The green dashed line is the contrast obtained with theoretical instantaneous control. The RL method provides a gain in contrast in particular in the direction of the dominant wind. Moreover, in low noise regime, RL provides raw contrast that is close to the theoretical limit of aliasing error.}
\label{fig:contrast_all}
\end{figure}

We compare the correction performance of the fully converged PETS models to the integrator in terms of raw point spread function (PSF) contrast \cite{perrin2003structure} and Strehl ratio \cite{roddier1999adaptive}. Each simulation run consists of 8000 frames, i.e. 16 sec. The resulting H-band Strehl ratios are computed from the wavefront error maps by Marechal's approximation \cite{roddier1999adaptive}, and are presented in Figure \ref{fig:strehl}. The MBRL control outperforms the integrator in all cases. A predictive capacity of the MBRL algorithm should result in an improved raw PSF contrast by reducing the notorious wind-driven halo (WDH) \cite{cantalloube2018origin}. The raw PSF contrast is given by the intensity ratio of the perfect coronagraphic PSF \cite{cavarroc2006fundamental} at a certain angular separation over to the peak intensity of the non-coronagraphic image. In Figure \ref{fig:contrast1}, we see that the RL method significantly reduces the WDH in all noise cases and hence delivers a better raw PSF contrast especially along the dominant wind direction. We also analyze a time series of one randomly picked actuator shown in Figure \ref{fig:timeseries}, and see that the RL method follows the non-delayed signal much closer than the integrator which exhibits the expected 2-frame delay between incident wavefront and correction by the DM.

\subsection{Performance under misregistration}\label{misreg}
Besides the predictive power, MBRL may provide other benefits for AO. One such benefit could be the automatic adaptation to dynamic MR between DM and WFS. MR is often introduced through mechanically or thermally induced flexure in a real AO system and negatively affects the performance if left uncompensated. Algorithms to detect and compensate for MR exist \cite{heritier2018new}, but combining these with a data-driven predictive control, might not be trivial or at least might need online tuning of hyper-parameters involved. In turn, RL does not make a specific assumption on the origin of error terms. Consequently, altogether the same algorithm with the same hyperparameters, including the reconstruction matrix C, also learns errors due to MR. Prospects are that RL might also learn to minimize some error terms we are not expecting.

\begin{figure}[ht]
\centering
    \includegraphics[trim={2.8cm 0 2cm 0},
    width=\textwidth]{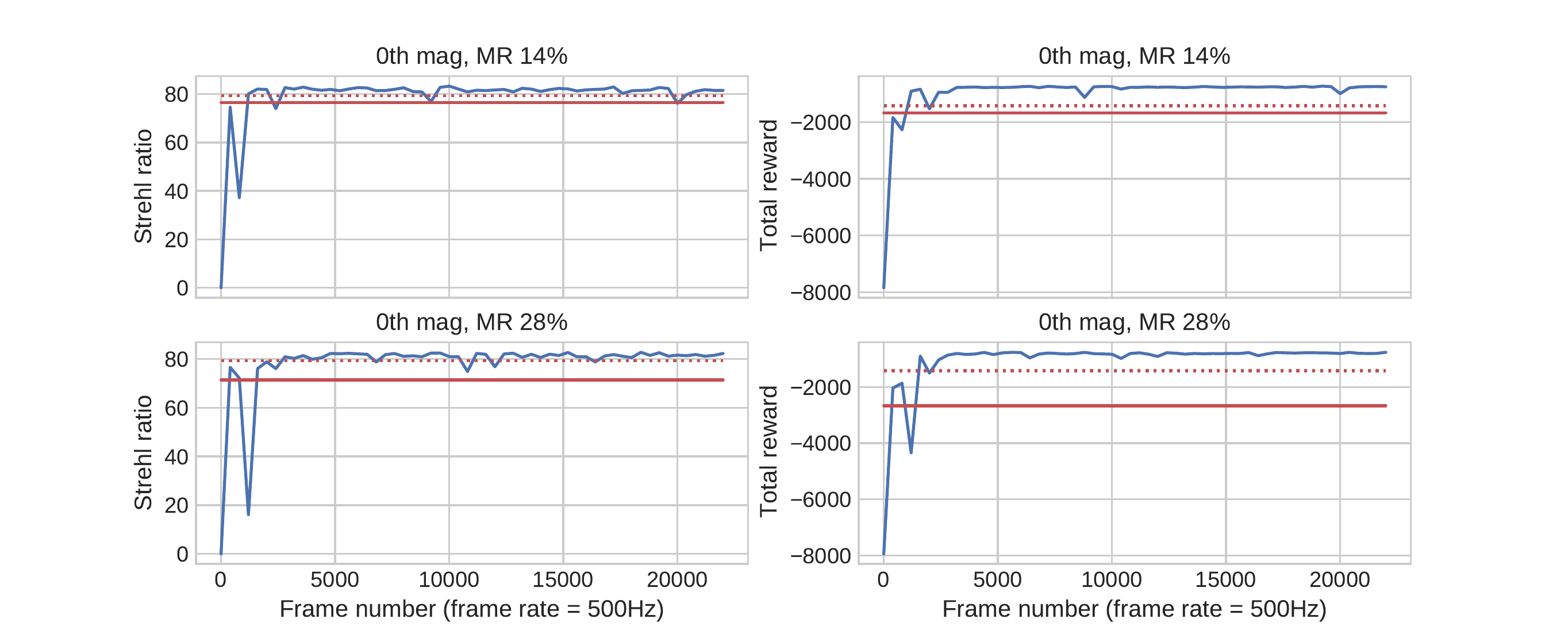}
    \caption{Learning curves under MR for proposed RL method. The performance vs time of the MBRL controller is shown in blue, and the mean Strehl ratio of the integrator is shown in red  (solid: subject to MR, dashed: no MR). Due to the non optimal geometry some high order modes are not visible in the WFS anymore and learning curves contain more variance. Nevertheless, in both cases, the RL algorithm reaches a better performance than the Intergrator with no MR.}
    \label{fig:results_training_mr}
\end{figure}

\begin{figure}[ht]
\centering
    \includegraphics[trim={4.4cm 0 3cm 0},clip, width=\textwidth]{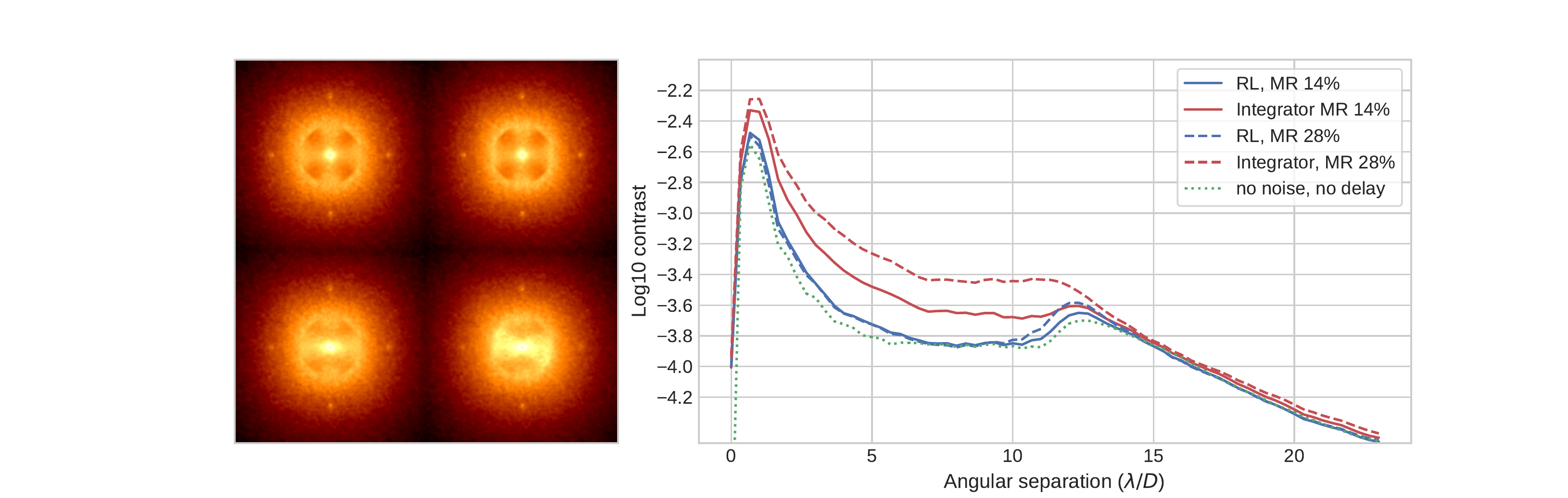}
    \caption{Predictive control under mis-registration. Left: Raw PSF contrast on the pupil plane. Right: The radial averages over the image. The blue lines are for PETS algorithm and red for the integrator with non corrected MR. The PETS maintains the performance under a sever MR.}
    \label{fig:contrast4}
\end{figure}

In order to verify this claim, we ran a simulation of the bright guide star case while shifting the WFS with respect to the DM by $14\%$ to the upper left (1 px up and 1 pix right on the WFS) and in another case by $28\%$ to the lower right (2 px down and 2 px right). Note that the reconstruction matrix C does not include the MR, i.e., the residual voltage presentation of WFS measurement does not match the mirror's voltage presentation anymore.  
The results are shown in Figures \ref{fig:results_training_mr} and \ref{fig:contrast4}. The MBRL control maintains its performance and predictive capacity even when a serve MR of 28\% of a subaperture is applied. Only at high spatial frequencies close to the DM correction radius \cite{perrin2003structure}, we see a small contrast degradation in the 28\% MR case. This is due to the non optimal alignment geometry, i.e., some higher order modes on the DM are not anymore visible in the WFS. The RL method also learns to stabilize these modes.

\section{Discussion}

We have formulated the control task of a closed-loop adaptive optics system as a Markov decision process and evaluated the performance of standard deep reinforcement learning algorithms on such a system. Our simulation results demonstrate that a state-of-the-art MBRL algorithm PETS robustly performs well with no environment-specific assumptions, apart from a generic reconstruction matrix. Moreover, the MBRL method predicted the turbulence evolution to a good approximation and automatically adapted to misregistration between DM and WFS, and was robust to measurement noise. Even though the algorithm itself is rather complicated to implement, its usage is simple: the algorithm calibrates, tunes, and maintains itself automatically.  

The MBRL method operates on control voltages and residual voltages which are derived from the residual WFS measurements and takes into account closed-loop dynamics along with the temporal evolution of the atmosphere. All the data needed is recorded on the control system itself eliminating dependencies on any numerical simulator or assumptions on the physics of the system. The MBRL control also outperformed classical integrator control in all simulation environments considered in Section \ref{sec:results}. The simulated performance is limited by the aliasing error of the SHS. With our single sensor setup and the objective to null future measurements, the correction of the DM unavoidably includes low spatial frequency aberrations which cancel the SHS signal of high spatial frequency turbulence \cite{veran1997estimation}. Finally, the MBRL method learns quickly requiring only 1600 timesteps in the simulator to surpass the baseline controller and converges at around 4000 timesteps. 

We simulate a relatively low order system with 24 actuators across the pupil. On the one hand, this keeps the execution times low with our moderate computational resources. On the other hand, the chosen system size is very relevant, because it simulates the size foreseen for the second AO stages currently planned or under development \cite{boccaletti2020sphere+, chazelas2020ristretto, kasper2020ghost} and to be added to already existing first AO stages.  While here we consider a single stage SCAO system, our method could be extended to control such 2nd-stage AO by including the first stage's voltages in the state as well. 

In future work we plan to extend the algorithm and comprehensively study a system with more complex DM dynamics, non-linear WFS such as the Pyramid WFS, saturations, alignment errors, turbulence boiling, and a cascaded AO system with a fast second stage. In particular, the future extreme AO systems on the upcoming generation of extremely large telescopes will control more than $10^4$ degrees of freedom; as such, scalability of the method shall be considered.

Future work should also address the challenges imposed by a variable turbulence. Understanding the trade-off between model complexity and fast training is essential for a successful implementation. Our MBRL method already learns continuously on a timescales of several seconds. Therefore, prospects are good that it is capable of automatically adjusting to changing conditions on timescales where atmosphere parameters typically change \cite{poyneer2009experimental}

Finally, we believe that, the biggest and most important challenge for a successful on-sky implementation of MBRL control for AO is the computational complexity of the method. In this work, the computational time at each timestep of the MPC on 448 degrees of freedom is around 80-120ms using a laptop equipped with a single NVIDIA Quadro RTX 3000 GPU and a straightforward implementation in PyTorch \cite{paszke2019pytorch}. Both, the delay and the temporal jitter are too large for a stable control of a real system.  In contrast to a real system whose cadence is defined by the atmosphere and WFS framerate, our simulations are stepwise and, therefore, not sensitive to the jitter, and no strategies to minimize it was devised. Jitter could, for example, be mitigated by exiting the planning algorithm after a given time rather than after a fixed number of iterations (20 in our simulations).

The large computational cost could be alleviated by reducing the number of parameters in the dynamics model, employing fewer samples in the planning phase, and tuning the CEM procedure's hyperparameters. It seems feasible that these points combined with better hardware and optimized low-level implementation are sufficient to bring the running time of our method with 448 degrees of freedom down into the range needed for an on-sky system.

However, the algorithm's brute force approach could possibly be improved. A promising approach to speed up the  MBRL control system, could be to replace the dynamics model and/or the planning algorithm to reduce computational complexity. We proposed a dynamics model composed of an ensemble of convolutional NNs. If the non-linear property of NNs turns out not to be needed, a much simpler linear model, e.g., an autoregressive model, could be used instead. Also, we are already investigating other methods that replace the planning phase of MBRL with a so-called policy function  \cite{deisenroth2011pilco}, which could be implemented as a NN and therefore avoid iterations and make the controller fast.

Finally, an efficient possible direction to reduce computational effort is to apply MBRL control only to a low-dimensional subset of the controlled parameters. For example, modal control could allow us to control a small set of modes with MBRL, while other modes are controlled classically.

\section*{Funding}
JN was supported by the European Southern Observatory. Moreover, JN and TH acknowledge support by the Academy of Finland via the Academy Research Fellow project, decision number 326961. \\

\section*{Acknowledgments} 
We thank Arto Klami, Cedric Heritier, Miska Le Louarn for fruitful discussions and guidance on the project. \\

\section*{Disclosures} 
The authors declare no conflicts of interest.

\bibliography{references}

\begin{thebibliography}{10}
\newcommand{\enquote}[1]{``#1''}

\bibitem{babcock1953possibility}
H.~W. Babcock, \enquote{The possibility of compensating astronomical seeing,}
  {\protect\JournalTitle{Publications of the Astronomical Society of the
  Pacific}} \textbf{65}, 229--236 (1953).

\bibitem{hardy1998adaptive}
J.~W. Hardy, \emph{Adaptive optics for astronomical telescopes}, vol.~16
  (Oxford University, 1998).

\bibitem{roddier1999adaptive}
F.~Roddier, \emph{Adaptive optics in astronomy} (Cambridge University, 1999).

\bibitem{guyon2005limits}
O.~Guyon, \enquote{Limits of adaptive optics for high-contrast imaging,}
  {\protect\JournalTitle{The Astrophysical Journal}} \textbf{629}, 592 (2005).

\bibitem{heritier2018new}
C.~Heritier, S.~Esposito, T.~Fusco, B.~Neichel, S.~Oberti, R.~Briguglio,
  G.~Agapito, A.~Puglisi, E.~Pinna, and P.~Y. Madec, \enquote{A new calibration
  strategy for adaptive telescopes with pyramid wfs,}
  {\protect\JournalTitle{Monthly Notices of the Royal Astronomical Society}}
  \textbf{481}, 2829--2840 (2018).

\bibitem{mnih2013playing}
V.~Mnih, K.~Kavukcuoglu, D.~Silver, A.~Graves, I.~Antonoglou, D.~Wierstra, and
  M.~Riedmiller, \enquote{Playing atari with deep reinforcement learning,}
  {\protect\JournalTitle{arXiv preprint arXiv:1312.5602}}  (2013).

\bibitem{silver2017mastering}
D.~Silver, T.~Hubert, J.~Schrittwieser, I.~Antonoglou, M.~Lai, A.~Guez,
  M.~Lanctot, L.~Sifre, D.~Kumaran, T.~Graepel, T.~P. Lillicrap, K.~Simonyan,
  and D.~Hassabis, \enquote{Mastering chess and shogi by self-play with a
  general reinforcement learning algorithm,} {\protect\JournalTitle{arXiv
  preprint arXiv:1712.01815}}  (2017).

\bibitem{zhang2015towards}
F.~Zhang, J.~Leitner, M.~Milford, B.~Upcroft, and P.~Corke, \enquote{Towards
  vision-based deep reinforcement learning for robotic motion control,}
  {\protect\JournalTitle{arXiv preprint arXiv:1511.03791}}  (2015).

\bibitem{kalashnikov2018qt}
D.~Kalashnikov, A.~Irpan, P.~Pastor, J.~Ibarz, A.~Herzog, E.~Jang, D.~Quillen,
  E.~Holly, M.~Kalakrishnan, V.~Vanhoucke, and S.~Levine, \enquote{Qt-opt:
  Scalable deep reinforcement learning for vision-based robotic manipulation,}
  {\protect\JournalTitle{arXiv preprint arXiv:1806.10293}}  (2018).

\bibitem{sutton2018reinforcement}
R.~S. Sutton and A.~G. Barto, \emph{Reinforcement learning: An introduction}
  (Massachusetts Institute of Technology, 2018).

\bibitem{chua2018deep}
K.~Chua, R.~Calandra, R.~McAllister, and S.~Levine, \enquote{Deep reinforcement
  learning in a handful of trials using probabilistic dynamics models,} in
  \emph{Advances in Neural Information Processing Systems,}  (2018), pp.
  4754--4765.

\bibitem{kulcsar2006optimal}
C.~Kulcs{\'a}r, H.-F. Raynaud, C.~Petit, J.-M. Conan, and P.~V.~D. Lesegno,
  \enquote{Optimal control, observers and integrators in adaptive optics,}
  {\protect\JournalTitle{Optics express, 14(17):7464--7476}}  (2006).

\bibitem{paschall1993linear}
R.~N. Paschall and D.~J. Anderson, \enquote{Linear quadratic gaussian control
  of a deformable mirror adaptive optics system with time-delayed
  measurements,} {\protect\JournalTitle{Applied optics}} \textbf{32},
  6347--6358 (1993).

\bibitem{gray2012ensemble}
M.~Gray and B.~Le~Roux, \enquote{Ensemble transform kalman filter, a
  nonstationary control law for complex ao systems on elts: theoretical aspects
  and first simulations results,} in \emph{Adaptive Optics Systems III,}  vol.
  8447 (International Society for Optics and Photonics, 2012), p. 84471T.

\bibitem{conan1a2011integral}
J.-M. Conan, H.~Raynaud, C.~AR, Kulcs{\'a}r, S.~Meimon, and G.~Sivo,
  \enquote{Are integral controllers adapted to the new era of elt adaptive
  optics?} in \emph{AO4ELT,}  (2011).

\bibitem{correia2010adapting}
C.~Correia, J.-M. Conan, C.~Kulcs{\'a}r, H.-F. Raynaud, and C.~Petit,
  \enquote{Adapting optimal lqg methods to elt-sized ao systems,} in \emph{1st
  AO4ELT conference-Adaptive Optics for Extremely Large Telescopes,}  (EDP
  Sciences, 2010), p. 07003.

\bibitem{correia2010optimal}
C.~Correia, H.-F. Raynaud, C.~Kulcs{\'a}r, and J.-M. Conan, \enquote{On the
  optimal reconstruction and control of adaptive optical systems with mirror
  dynamics,} {\protect\JournalTitle{JOSA A}} \textbf{27}, 333--349 (2010).

\bibitem{correia2017modeling}
C.~M. Correia, C.~Z. Bond, J.-F. Sauvage, T.~Fusco, R.~Conan, and P.~L.
  Wizinowich, \enquote{Modeling astronomical adaptive optics performance with
  temporally filtered wiener reconstruction of slope data,}
  {\protect\JournalTitle{JOSA A}} \textbf{34}, 1877--1887 (2017).

\bibitem{poyneer2007fourier}
L.~A. Poyneer, B.~A. Macintosh, and J.-P. V{\'e}ran, \enquote{Fourier transform
  wavefront control with adaptive prediction of the atmosphere,}
  {\protect\JournalTitle{JOSA A}} \textbf{24}, 2645--2660 (2007).

\bibitem{males2018ground}
J.~R. Males and O.~Guyon, \enquote{Ground-based adaptive optics coronagraphic
  performance under closed-loop predictive control,}
  {\protect\JournalTitle{Journal of Astronomical Telescopes, Instruments, and
  Systems}} \textbf{4}, 019001 (2018).

\bibitem{dessenne1998optimization}
C.~Dessenne, P.-Y. Madec, and G.~Rousset, \enquote{Optimization of a predictive
  controller for closed-loop adaptive optics,} {\protect\JournalTitle{Applied
  optics}} \textbf{37}, 4623--4633 (1998).

\bibitem{guyon2017adaptive}
O.~Guyon and J.~Males, \enquote{Adaptive optics predictive control with
  empirical orthogonal functions (eofs),} {\protect\JournalTitle{arXiv preprint
  arXiv:1707.00570}}  (2017).

\bibitem{van2017performance}
M.~van Kooten, N.~Doelman, and M.~Kenworthy, \emph{Performance of AO predictive
  control in the presence of non-stationary turbulence} (Instituto de
  Astrofisica de Canarias, 2017).

\bibitem{van2019impact}
M.~van Kooten, N.~Doelman, and M.~Kenworthy, \enquote{Impact of time-variant
  turbulence behavior on prediction for adaptive optics systems,}
  {\protect\JournalTitle{JOSA A}} \textbf{36}, 731--740 (2019).

\bibitem{haffert2021datadriven}
S.~Y. Haffert, J.~R. Males, L.~M. Close, K.~V. Gorkom, J.~D. Long, A.~D.
  Hedglen, O.~Guyon, L.~Schatz, M.~Kautz, J.~Lumbres, A.~Rodack, J.~M. Knight,
  H.~Sun, and K.~Fogarty, \enquote{Data-driven subspace predictive control of
  adaptive optics for high-contrast imaging,}  (2021).

\bibitem{swanson2018wavefront}
R.~Swanson, M.~Lamb, C.~Correia, S.~Sivanandam, and K.~Kutulakos,
  \enquote{Wavefront reconstruction and prediction with convolutional neural
  networks,} in \emph{Adaptive Optics Systems VI,}  vol. 10703 (International
  Society for Optics and Photonics, 2018), p. 107031F.

\bibitem{liu2019using}
X.~Liu, T.~Morris, and C.~Saunter, \enquote{Using long short-term memory for
  wavefront prediction in adaptive optics,} in \emph{International Conference
  on Artificial Neural Networks,}  (Springer, 2019), pp. 537--542.

\bibitem{sun2017bayesian}
Z.~Sun, Y.~Chen, X.~Li, X.~Qin, and H.~Wang, \enquote{A bayesian regularized
  artificial neural network for adaptive optics forecasting,}
  {\protect\JournalTitle{Optics Communications}} \textbf{382}, 519--527 (2017).

\bibitem{mcguire1999adaptive}
P.~C. McGuire, D.~G. Sandler, M.~Lloyd-Hart, and T.~A. Rhoadarmer,
  \enquote{Adaptive optics: Neural network wavefront sensing, reconstruction,
  and prediction,} in \emph{Scientific Applications of Neural Nets,}
  (Springer, 1999), pp. 97--138.

\bibitem{jensen2019demonstrating}
R.~Jensen-Clem, C.~Z. Bond, S.~Cetre, E.~McEwen, P.~Wizinowich, S.~Ragland,
  D.~Mawet, and J.~Graham, \enquote{Demonstrating predictive wavefront control
  with the keck ii near-infrared pyramid wavefront sensor,} in \emph{Techniques
  and Instrumentation for Detection of Exoplanets IX,}  vol. 11117
  (International Society for Optics and Photonics, 2019), p. 111170W.

\bibitem{gomez2019experience}
S.~L.~S. G{\'o}mez, C.~Gonz{\'a}lez-Guti{\'e}rrez, E.~D. Alonso, J.~D. Santos,
  M.~L.~S. Rodr{\'\i}guez, T.~Morris, J.~Osborn, A.~Basden, L.~Bonavera,
  J.~G.-N. Gonz{\'a}lez, and F.~J. de~Cos~Juez, \enquote{Experience with
  artificial neural networks applied in multi-object adaptive optics,}
  {\protect\JournalTitle{Publications of the Astronomical Society of the
  Pacific}} \textbf{131}, 108012 (2019).

\bibitem{osborn2014open}
J.~Osborn, D.~Guzman, F.~J. de~Cos~Juez, A.~G. Basden, T.~J. Morris,
  E.~Gendron, T.~Butterley, R.~M. Myers, A.~Guesalaga, F.~Sanchez~Lasheras,
  M.~Gomez~Victoria, M.~L. Sánchez~Rodríguez, D.~Gratadour, and G.~Rousset,
  \enquote{Open-loop tomography with artificial neural networks on canary:
  on-sky results,} {\protect\JournalTitle{Monthly Notices of the Royal
  Astronomical Society, 441(3):2508--2514}}  (2014).

\bibitem{gonzalez2017comparative}
C.~Gonz{\'a}lez-Guti{\'e}rrez, J.~D. Santos, M.~Mart{\'\i}nez-Zarzuela, A.~G.
  Basden, J.~Osborn, F.~J. D{\'\i}az-Pernas, and F.~J. de~Cos~Juez,
  \enquote{Comparative study of neural network frameworks for the next
  generation of adaptive optics systems,} {\protect\JournalTitle{Sensors}}
  \textbf{17}, 1263 (2017).

\bibitem{sandler1991use}
D.~Sandler, T.~Barrett, D.~Palmer, R.~Fugate, and W.~Wild, \enquote{Use of a
  neural network to control an adaptive optics system for an astronomical
  telescope,} {\protect\JournalTitle{Nature}} \textbf{351}, 300--302 (1991).

\bibitem{Landman:20}
R.~Landman and S.~Y. Haffert, \enquote{Nonlinear wavefront reconstruction with
  convolutional neural networks for fourier-based wavefront sensors,}
  {\protect\JournalTitle{Opt. Express}} \textbf{28}, 16644--16657 (2020).

\bibitem{xu2019deep}
Z.~Xu, P.~Yang, K.~Hu, B.~Xu, and H.~Li, \enquote{Deep learning control model
  for adaptive optics systems,} {\protect\JournalTitle{Applied optics}}
  \textbf{58}, 1998--2009 (2019).

\bibitem{ke2019self}
H.~Ke, B.~Xu, Z.~Xu, L.~Wen, P.~Yang, S.~Wang, and L.~Dong,
  \enquote{Self-learning control for wavefront sensorless adaptive optics
  system through deep reinforcement learning,} {\protect\JournalTitle{Optik}}
  \textbf{178}, 785--793 (2019).

\bibitem{hu2018build}
K.~Hu, Z.~Xu, W.~Yang, and B.~Xu, \enquote{Build the structure of wfsless ao
  system through deep reinforcement learning,} {\protect\JournalTitle{IEEE
  Photonics Technology Letters}} \textbf{30}, 2033--2036 (2018).

\bibitem{landman2020self}
R.~Landman, S.~Y. Haffert, V.~M. Radhakrishnan, and C.~U. Keller,
  \enquote{Self-optimizing adaptive optics control with reinforcement
  learning,} in \emph{Adaptive Optics Systems VII,}  vol. 11448 (International
  Society for Optics and Photonics, 2020), p. 1144849.

\bibitem{gendron1994astronomical}
E.~Gendron and P.~L{\'e}na, \enquote{Astronomical adaptive optics. 1: Modal
  control optimization,} {\protect\JournalTitle{Astronomy and Astrophysics}}
  \textbf{291}, 337--347 (1994).

\bibitem{madec1999control}
P.-Y. Madec, \enquote{Control techniques,} {\protect\JournalTitle{Adaptive
  optics in astronomy}} pp. 131--154 (1999).

\bibitem{poyneer2009experimental}
L.~Poyneer, M.~van Dam, and J.-P. V{\'e}ran, \enquote{Experimental verification
  of the frozen flow atmospheric turbulence assumption with use of astronomical
  adaptive optics telemetry,} {\protect\JournalTitle{JOSA A}} \textbf{26},
  833--846 (2009).

\bibitem{conan2014object}
R.~Conan and C.~Correia, \enquote{Object-oriented matlab adaptive optics
  toolbox,} in \emph{Adaptive optics systems IV,}  vol. 9148 (International
  Society for Optics and Photonics, 2014), p. 91486C.

\bibitem{camacho2013model}
E.~F. Camacho and C.~B. Alba, \emph{Model predictive control} (Springer Science
  \& Business Media, 2013).

\bibitem{efron1994introduction}
B.~Efron and R.~J. Tibshirani, \emph{An introduction to the bootstrap} (CRC
  press, 1994).

\bibitem{lakshminarayanan2016simple}
B.~Lakshminarayanan, A.~Pritzel, and C.~Blundell, \enquote{Simple and scalable
  predictive uncertainty estimation using deep ensembles,}
  {\protect\JournalTitle{arXiv preprint arXiv:1612.01474}}  (2016).

\bibitem{Maas2013RectifierNI}
A.~L. Maas, A.~Y. Hannun, and A.~Y. Ng, \enquote{Rectifier nonlinearities
  improve neural network acoustic models,} in \emph{Proc. icml,}  vol.~30
  (2013), p.~3.

\bibitem{kingma2014adam}
D.~P. Kingma and J.~Ba, \enquote{Adam: A method for stochastic optimization,}
  {\protect\JournalTitle{arXiv preprint arXiv:1412.6980}}  (2014).

\bibitem{rigaut1998analytical}
F.~J. Rigaut, J.-P. Veran, and O.~Lai, \enquote{{Analytical model for
  Shack-Hartmann-based adaptive optics systems},} in \emph{Adaptive Optical
  System Technologies,}  vol. 3353 D.~Bonaccini and R.~K. Tyson, eds.,
  International Society for Optics and Photonics (SPIE, 1998), pp. 1038 --
  1048.

\bibitem{perrin2003structure}
M.~D. Perrin, A.~Sivaramakrishnan, R.~B. Makidon, B.~R. Oppenheimer, and J.~R.
  Graham, \enquote{The structure of high strehl ratio point-spread functions,}
  {\protect\JournalTitle{The Astrophysical Journal}} \textbf{596}, 702 (2003).

\bibitem{cantalloube2018origin}
F.~Cantalloube, E.~H. Por, K.~Dohlen, J.-F. Sauvage, A.~Vigan, M.~Kasper,
  N.~Bharmal, T.~Henning, W.~Brandner, J.~Milli, C.~Correia, and T.~Fusco,
  \enquote{Origin of the asymmetry of the wind driven halo observed in
  high-contrast images,} {\protect\JournalTitle{Astronomy \& Astrophysics}}
  \textbf{620}, L10 (2018).

\bibitem{cavarroc2006fundamental}
C.~Cavarroc, A.~Boccaletti, P.~Baudoz, T.~Fusco, and D.~Rouan,
  \enquote{Fundamental limitations on earth-like planet detection with
  extremely large telescopes,} {\protect\JournalTitle{Astronomy \&
  Astrophysics}} \textbf{447}, 397--403 (2006).

\bibitem{veran1997estimation}
J.-P. Veran, F.~J. Rigaut, H.~Maitre, and D.~Rouan, \enquote{Estimation of the
  adaptive optics long-exposure point spread function using control loop data:
  recent developments,} in \emph{Adaptive Optics and Applications,}  vol. 3126
  (International Society for Optics and Photonics, 1997), pp. 81--92.

\bibitem{boccaletti2020sphere+}
A.~Boccaletti, G.~Chauvin, D.~Mouillet, O.~Absil, F.~Allard, S.~Antoniucci,
  J.-C. Augereau, P.~Barge, A.~Baruffolo, J.-L. Baudino, P.~Baudoz,
  M.~Beaulieu, M.~Benisty, J.-L. Beuzit, A.~Bianco, B.~Biller, B.~Bonavita,
  M.~Bonnefoy, S.~Bos, J.-C. Bouret, W.~Brandner, N.~Buchschache, B.~Carry,
  F.~Cantalloube, E.~Cascone, A.~Carlotti, B.~Charnay, A.~Chiavassa,
  E.~Choquet, Y.~Clenet, A.~Crida, J.~De~Boer, V.~De~Caprio, S.~Desidera, J.-M.
  Desert, J.-B. Delisle, P.~Delorme, K.~Dohlen, D.~Doelman, C.~Dominik,
  V.~Orazi, C.~Dougados, S.~Doute, D.~Fedele, M.~Feldt, F.~Ferreira,
  C.~Fontanive, T.~Fusco, R.~Galicher, A.~Garufi, E.~Gendron, A.~Ghedina,
  C.~Ginski, J.-F. Gonzalez, D.~Gratadour, R.~Gratton, T.~Guillot, S.~Haffert,
  J.~Hagelberg, T.~Henning, E.~Huby, M.~Janson, I.~Kamp, C.~Keller,
  M.~Kenworthy, P.~Kervella, Q.~Kral, J.~Kuhn, E.~Lagadec, G.~Laibe,
  M.~Langlois, A.-M. Lagrange, R.~Launhardt, L.~Leboulleux, H.~Le~Coroller,
  G.~Li~Causi, M.~Loupias, A.~Maire, G.~Marleau, F.~Martinache, P.~Martinez,
  D.~Mary, M.~Mattioli, J.~Mazoyer, H.~Meheut, F.~Menard, D.~Mesa, N.~Meunier,
  Y.~Miguel, J.~Milli, M.~Min, P.~Molliere, C.~Mordasini, G.~Moretto,
  L.~Mugnier, G.~Muro~Arena, N.~Nardetto, M.~N. Diaye, N.~Nesvadba,
  F.~Pedichini, P.~Pinilla, E.~Por, A.~Potier, S.~Quanz, J.~Rameau,
  R.~Roelfsema, D.~Rouan, E.~Rigliaco, B.~Salasnich, M.~Samland, J.-F. Sauvage,
  H.-M. Schmid, D.~Segransan, I.~Snellen, F.~Snik, F.~Soulez, E.~Stadler,
  D.~Stam, M.~Tallon, P.~Thebault, E.~Thiebaut, C.~Tschudi, S.~Udry, R.~van
  Holstein, P.~Vernazza, F.~Vidal, A.~Vigan, R.~Waters, F.~Wildi, M.~Willson,
  A.~Zanutta, A.~Zavagno, and A.~Zurlo, \enquote{Sphere+: Imaging young
  jupiters down to the snowline,} {\protect\JournalTitle{arXiv preprint
  arXiv:2003.05714}}  (2020).

\bibitem{chazelas2020ristretto}
B.~Chazelas, C.~Lovis, N.~Blind, J.~K{\"u}hn, L.~Genolet, I.~Hughes, M.~Turbet,
  J.~Hagelberg, N.~Restori, M.~Kasper, and N.~N.~C. Urra, \enquote{Ristretto: a
  pathfinder instrument for exoplanet atmosphere characterization,} in
  \emph{Adaptive Optics Systems VII,}  vol. 11448 (International Society for
  Optics and Photonics, 2020), p. 1144875.

\bibitem{kasper2020ghost}
M.~Kasper, N.~C. Urra, P.~Pathak, M.~Bonse, J.~Nousiainen, B.~Engler, C.~T.
  Heritier, J.~Kammerer, S.~Leveratto, C.~Rajani, P.~Bristow, M.~Le~Louarn,
  P.-Y. Madec, S.~Ströbele, C.~Verinaud, A.~Glauser, S.~P. Quanz, T.~Helin,
  C.~Keller, F.~Snik, A.~Boccaletti, G.~Chauvin, D.~Mouillet, C.~Kulcs{\'a}r,
  and H.-F. Raynaud, \enquote{Pcs – roadmap for exoearth imaging with the
  elt,} {\protect\JournalTitle{ESO Messenger}} \textbf{182} (2020).

\bibitem{paszke2019pytorch}
A.~Paszke, S.~Gross, F.~Massa, A.~Lerer, J.~Bradbury, G.~Chanan, T.~Killeen,
  Z.~Lin, N.~Gimelshein, L.~Antiga \emph{et~al.}, \enquote{Pytorch: An
  imperative style, high-performance deep learning library,} in \emph{Advances
  in Neural Information Processing Systems,}  (2019), pp. 8024--8035.

\bibitem{deisenroth2011pilco}
M.~Deisenroth and C.~E. Rasmussen, \enquote{Pilco: A model-based and
  data-efficient approach to policy search,} in \emph{Proceedings of the 28th
  International Conference on machine learning (ICML-11),}  (Citeseer, 2011),
  pp. 465--472.

\end{thebibliography}

\end{document}